\documentclass[aps,prb,preprint,groupedaddress,showpacs]{revtex4}

\usepackage{graphicx}
\usepackage{dcolumn} 
\usepackage{bm}      
\usepackage{amsmath} 

\newcommand{\fm}{{\phantom{-}}}

\begin{document}

\title{Electronic structure of charge-ordered Fe$_3$O$_4$ from calculated
optical, megneto-optical Kerr effect, and O K-edge x-ray absorption spectra}

\author{I.~Leonov$^{1}$\email[e-mail: ]{Ivan.Leonov@Physik.uni-Augsburg.de},
A.~N.~Yaresko$^2$, V.~N.~Antonov$^3$, and V.~I.~Anisimov$^4$}

\affiliation{$^1$ Theoretical Physics III, Center for Electronic Correlations and Magnetism, 
Institute for Physics, University of Augsburg, Germany}
\affiliation{$^2$ Max-Planck Institute for the Physics of Complex Systems, Dresden, Germany}
\affiliation{$^3$ Institute of Metal Physics, Vernadskii Street, 03142 Kiev, Ukraine}
\affiliation{$^4$ Institute of Metal Physics, Russian Academy of Science-Ural Division, 620219 
Yekaterinburg GSP-170, Russia}

\date{\today}

\begin{abstract}
The electronic structure of the low-temperature (LT) monoclinic magnetite,
Fe$_3$O$_4$, is investigated using the local spin density approximation
(LSDA) and the LSDA+$U$ method.  The self-consistent charge ordered
LSDA+$U$ solution has a pronounced $[001]$ charge density wave character.
In addition, a minor $[00\frac{1}{2}]$ modulation in the phase of the
charge order (CO) also occurs. While the existence of CO is evidenced by
the large difference between the occupancies of the minority spin $t_{2g}$
states of ``$2+$'' and ``$3+$'' Fe$_B$ cations, the total $3d$ charge
disproportion is small, in accord with the valence-bond-sum analysis of
structural data. Weak Fe orbital moments of $\sim$0.07$\mu_{\mathrm{B}}$
are obtained from relativistic calculations for the CO phase which is in
good agreement with recent x-ray magnetic circular dichroism measurements.
Optical, magneto-optical Kerr effect, and O $K$-edge x-ray absorption
spectra calculated for the charge ordered LSDA+$U$ solution are compared to
corresponding LSDA spectra and to available experimental data. The
reasonably good agreement between the theoretical and experimental spectra
supports the relevance of the CO solution obtained for the monoclinic LT
phase. The results of calculations of effective exchange coupling constants
between Fe spin magnetic moments are also presented.
\end{abstract}

\pacs{71.20.-b, 71.28.+d, 71.30.+h}

\maketitle

\section{Introduction}
\label{sec:int}

The problem of a theoretical description of metal-insulator transitions has a
challenging history of almost 70 years. It was first addressed by Verwey,
de Boer, and Peierls in the late 1930's; they pointed out the extremely
important role of electron-electron correlations in a partially filled
$d$-electron band in transition metal oxides,\cite{BV+37,P+37} e.g., nickel
oxide (NiO) and magnetite (Fe$_3$O$_4$). In both systems the
metal-insulator transition occurs, violating the Bloch-Wilson
band-insulator concept, the only one known at that
time.\cite{B29,S28,W31a,W31b} These earlier observations launched the long
and continuing history of the field of strongly correlated electrons. In
the past 70 years, much progress has been achieved from both
theoretical and experimental sides in understanding strongly correlated
electrons and metal-insulator transitions.\cite{IFT+98} However, the
charge ordering proposed by Verwey behind the metal-insulator transition
\cite{V39,VH41,VHR47} in 
Fe$_3$O$_4$ remains at the focus of active
debate.\cite{WAR01,SGB04}

Magnetite is a permanent natural magnet. Its magnetic properties have
fascinated mankind for several thousand years already.\cite{DCM+98}
Comprehensive investigation of magnetite started in 1913 when the first
indication of the first-order transition was obtained by a susceptibility
measurement on a synthetic polycrystal.\cite{R13, WR14} An anomalous
transition peak at $\sim$130 K was found in addition to the already known
paramagnetic-to-ferrimagnetic order transition on cooling through $\sim$850 K.
Electric resistivity measurements initially performed by Okamura\cite{O32}
and Verwey\cite{V39} showed that this anomalous behavior of the susceptibility
coincides with a sharp two orders of magnitude increase of resistivity on
cooling below $\sim$120 K.\cite{O32,ET34} The formal valency of Fe$_B$
cations on the octahedral $B$ sublattice of the inverted spinel AB$_2$O$_4$
structure is noninteger (2.5+). Verwey suggested that this transition is
caused by the ordering of Fe$_B$ cations, with a simple charge arrangement
of $(001)$ planes alternately occupied by $2+$ and $3+$ Fe$_B$
cations.\cite{V39,VH41,VHR47} This particular (Verwey) charge order (CO)
obeys the so-called Anderson criterion\cite{And56} of minimal electrostatic
repulsion leading to a short range CO pattern with each tetrahedron formed
by $B$ sites being occupied by equal numbers of 2+ and 3+ cations.

However, further experiments disproved the orthorhombic Verwey CO model.
These experiments have clearly established the rhombohedral distortions of
the cubic unit cell first detected by Rooksby $et~al.$ from x-ray powder
diffraction.\cite{TR51,RW53} Furthermore, observations of superstructure
reflections revealed half-integer satellite reflections, indexed as
$(h,k,l+\frac{1}{2})$ on the cubic unit cell, which points to a doubling of
the unit cell along the $c$ axis and shows the symmetry to be
monoclinic.\cite{SBD68,YSC68} The observation of monoclinic lattice
symmetry was also confirmed by a single-crystal x-ray study,\cite{YI79}
whereas the observation of a magnetoelectric effect indicated even lower
$P1$ symmetry in the low-temperature phase.\cite{MS93} Although, clear
evidence of the monoclinic lattice symmetry below $T_V$ was obtained, small
atomic displacements have not been fully resolved so far. The absence of a
definitive, experimentally determined structure gives rise to many
theoretical models proposed for the low-temperature (LT) phase of
magnetite.\cite{Rev01} In particular, purely electronic
\cite{CC71,CC73,Miz78,ZSP+90} and electron-phonon \cite{Mot80,Y80,IL86}
models for CO, as well as a bond dimerized ground state without charge
separation,\cite{SOF+02} have been proposed.

At room temperature Fe$_3$O$_4$ is a poor metal with an electronic
resistivity of 4 m$\Omega$ cm which is considerably higher than the resistivity
of simple metals (for instance, the resistivity of Cu is 1.7 $\mu \Omega$
cm).  The microscopic origin of the electronic state above the Verwey
transition is still unclear. Photoemission studies performed by Chainani
{\it et al.}\cite{CYM95} and Park {\it et al.}\cite{PTA97} gave
controversial results indicating metallic and semiconducting behavior above
$T_V$, respectively. In accordance with Park {\it et al.},\cite{PTA97}
qualitatively similar conclusions have been derived by Schrupp {\it et
al.}\cite{SST05} and Pimenov {\it et al.}\cite{PTR05} basing on a soft
x-ray photoemission study and complex conductivity measurements at
terahertz/infrared frequencies, respectively.

Magnetite is a ferrimagnet with anomalous high Curie temperature of
$\sim$850 K. The $A$-site magnetic moments are aligned antiparallel to the
$B$-site moments. This remarkable situation of antiferromagnetic coupling
between Fe$_A^{3+}$ and Fe$_B^{2.5+}$ cations, which have $d^5$ and
$d^{5.5}$ electron configurations in the high-spin state, respectively, should
lead to a half-metallic state with integer magnetic moment of 4$\mu_B$
per formula unit. This is in qualitatively good agreement with
magnetization measurements, which result in a $\sim$4.1\,$\mu_B$ magnetic
moment per unit cell.\cite{WF29,RS78} In contrast, spin resolved
photoemission measurements show $\sim$20\% suppression of the pure negative
spin polarization from the half-metallic behavior obtained for
Fe$_3$O$_4$ (111) films.\cite{DRG+02} Scanning tunneling microscopy and
spin-polarized photoelectron spectroscopy give only 55\% of negative spin
polarization at the Fermi level.\cite{FPD+05} To explain this phenomenon
two possible scenarios have been recently proposed. According to the first,
Fe$_3$O$_4$ is still a half metal in the bulk, whereas surface stresses
lead to unusual lattice distortions which reduce spin polarization at
the Fermi level to $\sim$40\%.\cite{PWM+05,FPD+05} An alternative
explanation is the formation of nonquasiparticle or incoherent states in
the gap near the Fermi level due to correlation effects.\cite{IKL+04}

Recent bond-valence-sum analysis \cite{BOK91} of high-resolution neutron
and x-ray powder diffraction data results in a small charge disproportion of
only 0.2\,$\bar{e}$ between Fe$_B$ cations with the 2+ and 3+
formal valency.\cite{WAR01,WAR02} This interpretation has been the
subject of much controversy.\cite{GSBP,SGB04} However, the smallness of
the charge order parameter was reproduced in an electronic structure study of
the refined low-temperature crystal structure using the loacl spin density
approximation (LSDA)+$U$
method.\cite{LYA04,JGH04} 
In particular, a more complicated charge-ordering
pattern inconsistent with the Verwey CO model was obtained.  In addition to
that, the $t_{2g}$ occupancy self-consistently obtained in the LSDA+$U$
calculations is strongly modulated between the Fe$^{2+}_{B}$ and
Fe$^{3+}_{B}$ cations, yielding a distinct orbital order with an order
parameter that reaches 70\% of the ideal value.\cite{LYA04} Since no direct
experimental confirmation of this charge and orbital order pattern is so
far available, the interpretation of these results is still open to
debate. However, this behavior seems to be universal and has recently been
found in several other charge ordered mixed-valent
systems.\cite{LYA+05,LYA+05b,JGH04,YLF06,L06}

In order to further check the pertinence of the CO model obtained
self-consistently in Ref.\ \onlinecite{LYA04} we carried out a detailed
theoretical study of exchange coupling constants, optical conductivity,
magneto-optical (MO) Kerr effect, and x-ray absorption at the O $K$ edge of
low-temperature Fe$_3$O$_4$ and compared the results of the calculations to
the available experimental data. 

The paper is organized as follows. In Sec.\ \ref{sec:struc} we discuss the
low temperature crystal structure of Fe$_3$O$_4$. Section \ref{sec:compd}
presents computational details relevant to reproduce the calculation
results. The results of electronic structure calculations obtained by the
LSDA and LSDA+$U$ methods for the low-temperature Fe$_3$O$_4$ are presented
in Secs.\ \ref{sec:lsda} and \ref{sec:lsdau}, respectively.  In Section
\ref{sec:lsdau} a charge and orbital order are also discussed and results
of calculations of exchange coupling constants are presented.  In Section
\ref{sec:optics} we present calculated optical, magneto-optical, and O
$K$-edge x-ray absorption spectra and compare them to corresponding
experimental spectra.  Finally, the results are summarized in Section
\ref{sec:sum}.

\section{Crystal structure and charge order}
\label{sec:struc}

Above the Verwey transition magnetite crystallizes in the face-centered
cubic (fcc) inverse spinel crystal structure with space group
$Fd\bar{3}m$. The iron atoms occupy the interstitial positions of the
close-packed fcc structure formed by the oxygen atoms. According to the
chemical formula of inverse spinel, $AB_2$O$_4$, there are two
different iron sublattices ($A$ and $B$) distinguished by the point
symmetry ($T_d$ and $D_{3d}$) as well as by the averaged Fe--O distance
(1.876 and 2.066 \AA\ for $A$ and $B$ iron sites, respectively) and the
valence state of iron cations.  In particular, the $A$ sublattice is formed
by Fe$^{3+}$ cations tetrahedrally coordinated by four oxygen ions, whereas
the so-called $B$ sublattice consists of iron sites octahedrally
coordinated by six oxygen ions. The octahedral $B$-sites are occupied by an
equal number of randomly distributed 2+ and 3+ Fe cations, which results in
an average valence value of 2.5+ per each Fe$_B$ cation.  The
$B$-sublattice is highly frustrated and can be considered as a diamond
lattice of Fe$_B$ cation tetrahedra, sharing corners with each other.

One of the driving forces for formation of the charge-ordered state in mixed
valent transition metal oxides is Coulomb repulsion.  In particular,
another key to understanding the charge-ordered structure in Fe$_3$O$_4$ is
provided by the Anderson criterion for the minimum of electrostatic
repulsion energy.\cite{And56} 
According to Anderson each tetrahedron formed by
octahedral $B$ sites of the spinel structure is occupied by two 2+ and two
3+ Fe cations in order to minimize the intersite electrostatic repulsion.
Thus, within the $Cc$ supercell of the cubic Fe$_3$O$_4$, there are ten
independent cationic arrangements, whereas one of them coincides with the
Verwey model.\cite{ZSP+90} However, the Verwey CO pattern has overall the
lowest classical electron correlation energy.

The low-temperature structure was shown to have a $\sqrt{2}a \times
\sqrt{2}a \times 2a$ supercell with space group $Cc$ from x-ray and neutron
diffraction.\cite{WAR01,WAR02} However, recent structural refinement (at
90 K) was only possible in the centric monoclinic space group $P2/c$ with
$\frac{a}{\sqrt{2}} \times \frac{a}{\sqrt{2}} \times 2a$ of the cubic
spinel subcell and eight formula units in the primitive unit
cell.\cite{WAR01,WAR02} Since the refinement for the $P2/c$ space group was
found to be unstable, additional $Pmca$ orthorhombic symmetry constraints
were also applied. This is equivalent to averaging the true superstructure
over the additional symmetry operators, i.e., each $B$ site in the $P2/c$ unit
cell is averaged over four nonequivalent subsites in the large $\sqrt{2}a
\times \sqrt{2}a \times 2a$\ $Cc$ supercell. Note, however, that such an
approximation is robust in the sense of smallness of any distortions from
the $P2/c$ subcell to the $Cc$ monoclinic cell (according to
Ref.~\onlinecite{WAR02} these are of $\sim$0.01 \AA). Previous structure
refinement below $T_V$ obtained by Iizumi {\it et al.} resulted in a
$\frac{a}{\sqrt{2}} \times \frac{a}{\sqrt{2}} \times 2a$ subcell of the
$Cc$ unit cell and imposed orthorhombic symmetry constraints on the atomic
positions.\cite{IKS82} In particular, a refinement based on an
approximation of the true crystal structure by a centric space group $Pmca$
or polar $Pmc2_1$ was proposed. But a charge-ordered arrangement has not
been identified in this refinement, although large atomic displacements of
Fe and O atoms were found. This is in strong qualitative contrast to the
recent structure refinement proposed by Wright {\it et al.}  where
clear evidence of CO below the transition has been found.\cite{WAR01,WAR02}

According to the refinement the octahedral Fe$_B$ sites are split into two
groups with different values of the averaged Fe--O bond distances; with
$B2$ and $B3$ sites being significantly smaller than $B1$ and $B4$
($B1$-$B4$ are crystallographically independent Fe$_B$ sites according to
the notation in Refs.~\onlinecite{WAR01} and \onlinecite{WAR02}).  A
different averaged Fe--O bond distance is a sensitive experimental
indicator of the cation charge state. Quantitative analysis of the valence
state of both Fe$_B$ groups using the bond-valence-sum (BVS) method shows
that the octahedral Fe$_B$ sites fall into two clear groups with respect to
the estimated value of valence. The result is a charge disproportion of
0.2\,$\bar{e}$ between large ($B1$ and $B4$) and small ($B2$ and $B3$)
sites (which has been referred as the class I CO model). Another possible
class of CO arises from the symmetry-averaging orthorhombic constraint.
There are 32 charge-ordered models which are refered to as class II CO
because large ($B1$ and $B4$) and small ($B2$ and $B3$) sites could be
averaged over (3 Fe$^{2+}$ + Fe$^{3+}$) and (Fe$^{2+}$ + 3 Fe$^{2+}$)
subsites, respectively.  The symmetry averaging results in decrease of the
more pronounced charge separation of 0.4$\bar{e}$ in the full $Cc$
superstructure (class II CO) down to 0.2\,$\bar{e}$ in the $P2/c$ subcell.
The Anderson criterion is not satisfied by any of the class I or class II
CO models. This is remarkable because the Anderson criterion has been
widely used in many CO models.\cite{Miz78,ZSP+90} However, class II, as was
shown from electrostatic repulsion energy estimations, appears to be more
plausible than the class I arrangement.

Recently this interpretation of the refined crystal structure has been
found to be controversial. The lack of atomic long-range CO and, as a
result, an intermediate valence regime below the Verwey transition were
proposed.\cite{GSBP,SGB04} It is argued that the difference of the average
Fe-O distances between compressed and expanded FeO$_6$ octahedra, which
could be considered as a maximum limit of charge disproportionation, has
the same order as the total sensitivity (including experimental errors) of
the bond-valence-sum method.  This remarkable controversy shows that the
understanding of the system is far from satisfactory.

\section{Computational details}
\label{sec:compd}

In this paper we present results of band structure calculations which
have been performed for the low-temperature crystal
structure of Fe$_3$O$_4$ recently refined at 90 K.\cite{WAR01,WAR02} The
refined cell parameters 
used in calculations are $a= 5.94437$~\AA, $b= 5.92471$~\AA, $c=
16.77512$~\AA, and $\beta= 90.236^{\circ}$. The monoclinic $P2/c$ unit cell
contains eight formula units. There are eight different types of iron atom
sites: two tetrahedrally coordinated $A$ sites and six $B$ sites
octahedrally coordinated by six O atoms.

The electronic structure of magnetite was calculated self-consistently
using the local-spin-density approximation and LSDA+$U$ approach
\cite{ldau:AZA91,ldau:LAZ95} with the linear muffin-tin orbitals (LMTO)
method in the atomic-sphere approximation.\cite{lda:OKA75} The radii of the
muffin-tin spheres were taken as $R_{\rm Fe}$= 2.125 a.u. and $R_{\rm O}$=
2.0 a.u. Fifteen kinds of empty spheres were introduced to fill up the
interatomic space.  As was shown in Ref.~\onlinecite{AHA+01}, weak
spin-orbit coupling does not appreciably affect the band structure of
Fe$_3$O$_4$. We neglect it for simplicity in Secs.\ \ref{sec:lsda} and
\ref{sec:lsdau}.  Optical, magneto-optical (MO), and x-ray absorption
spectra (XAS) for the $P2/c$ model of the LT phase of Fe$_3$O$_4$ were
calculated taking into account the spin-orbit coupling using the
spin-polarized relativistic LMTO method. \cite{PYLMTO} In order to
calculate O $K$-edge XAS spectra and the absorption parts of the optical
conductivity in a wide energy range Fe $4f$ and O $3d$ states were included
into the LMTO basis set. The dispersive parts of the conductivity were
calculated using the Kramers-Kronig relations.  XAS spectra were calculated
in the dipole approximation neglecting relaxation effects caused by the O
$1s$ core hole.
A detailed description of the formalism for the calculations of MO and
x-ray absorption spectra and corresponding matrix elements using the
relativistic LMTO method can be found in Refs.\
\onlinecite{book:AHY04,AHA+01,AHY+03} and references therein.

\section{LSDA band structure}
\label{sec:lsda}

\begin{figure}[tbp!]
\centerline{\includegraphics[width=0.45\textwidth,clip]{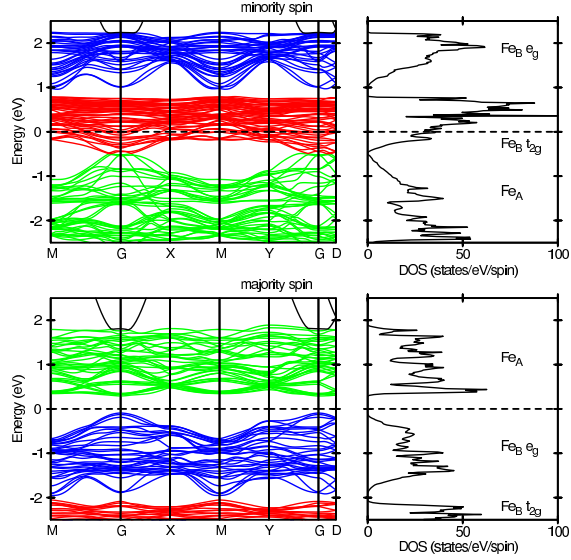}}
\caption{\label{fig:lsdabnd} (Color online)
Total density of states and band structure of the $P2/c$ phase of
Fe$_3$O$_4$ self-consistently obtained using the LSDA method. The Fermi
level is shown by dotted lines. The energy bands predominantly originating
from the Fe$_A$ $3d$ states are shown in light gray (green) color, whereas
Fe$_B$ $t_{2g}$ and $e_g$ bands are in dark gray (red) and gray (blue),
respectively. The corresponding contributions to the total DOS are shown in
the right panel.}
\end{figure}

Figure~\ref{fig:lsdabnd} shows the total density of states (DOS) and band
structure of Fe$_3$O$_4$ obtained from the LSDA
calculations. Interestingly, the crystal structure distortion taken
explicitly in the monoclinic phase does not strongly affect the electronic
structure of Fe$_3$O$_4$. In particular, the calculation results agree well
with previous band structure calculations for the cubic
phase.\cite{ZS91,YH99} Thus, the LSDA gives a uniform half-metallic
ferrimagnetic solution with partially filled bands originating from the
minority spin $t_{2g}$ orbitals of Fe$_B$ cations. An energy gap of
$\sim$1 eV opens in the majority spin channel between occupied Fe$_B$ $e_g$
and the bottom of the empty Fe$_A$ states.  The lower part of the valence band
(below $-$3.5 eV) is mainly formed by O $2p$ states with a bonding
hybridization with Fe $3d$ states, whereas bands near the Fermi level,
between $-$3.5 and 2 eV, have a predominant contribution of Fe $3d$
states.

It is worth recalling that Fe$_B$ 3$d$ states are split by the cubic
component of a ligand field into a triplet $t_{2g}$ and a doublet $e_{g}$.
Already in the $Fd\bar{3}m$ phase the local symmetry of Fe$_B$ sites
($D3d$) is lower than cubic and the trigonal component of the ligand field
splits $t_{2g}$ states into a singlet $a_{1g}$ and doublet $e'_g$. In the
monoclinic LT phase the symmetry is further lowered by the distortions and
all the degeneracy is lifted. Nevertheless, the cubic component of the
ligand field, which is determined by the relative strength of Fe $3d$--O
$2p$ hybridization of $\pi$ and $\sigma$ type, remains dominant, whereas
the splitting within $t_{2g}$ and $e_{g}$ subbands is smaller than the
corresponding bandwidth. This allows one to label the corresponding states
as $t_{2g}$ and $e_{g}$. Note, however, that the $P2/c$ frame is rotated by
$\sim\frac{\pi}{4}$ with respect to the $Fd\bar{3}m$ one and the angular
dependence of the $t_{2g}$ states is given by $d_{xz}\pm d_{yz}$ and
$d_{x^2-y^2}$ combinations of cubic harmonics.

\begin{figure}[tbp!]
\centerline{\includegraphics[width=0.43\textwidth,clip]{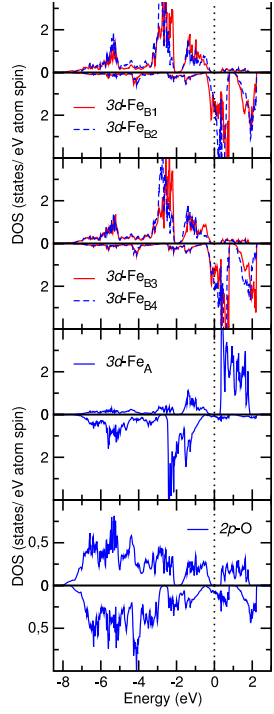}}
\caption{\label{fig:lsdapdos} (Color online) 
Partial DOS obtained from the LSDA calculations for the $P2/c$ phase of 
Fe$_3$O$_4$. The Fermi level is shown by dotted lines.}
\end{figure}

Figure~\ref{fig:lsdapdos} presents different contributions to the 
total density of states given by independent iron and oxygen sites. 
Crystallographically independent iron sites are labeled according to the 
notation of Ref.~\onlinecite{WAR01}. Note, however, that the DOSs for two 
pairs of sites that were constrained to have equivalent coordination during 
the structural refinement -- Fe$_{B1a}$, Fe$_{B1b}$ and 
Fe$_{B2a}$, Fe$_{B2b}$ -- are very similar and we present the DOS averaged 
over the corresponding pairs, dropping $a$ and $b$ indices.

The splitting of Fe$_B$ states due to the octahedral component of the
ligand field is roughly of 1 eV.  This is considerably smaller than the
exchange splitting between minority and majority spin states, which is
$\sim$3 eV, and consistent with the high-spin state of Fe cations. The
absolute values of magnetic moments obtained by LSDA are 3.14\,$\mu_B$,
3.30\,$\mu_B$, 3.44\,$\mu_B$, 3.27\,$\mu_B$, and 3.38\,$\mu_B$ for Fe$_A$
and Fe$_{B1-B4}$ cations, respectively. The total ferromagnetic moment per
Fe$_3$O$_4$ formula unit is 4.0\,$\mu_B$.

It should be noted that in contrast to experimental data the LSDA
calculations result in a half-metallic solution without charge ordering.
Apparently, the change of the LSDA electronic structure, produced by the
crystal structure distortion from the cubic to the monoclinic phase, is not
sufficient to explain the charge ordering and metal-insulator transition in
Fe$_3$O$_4$. The electron-electron correlations, mainly in the $3d$ shell
of Fe cations, play a significant role.

\section{LSDA+$U$ results}
\label{sec:lsdau}

\subsection{Band structure}
\label{sec:lsdabnd}

In order to account for the strong electronic correlations in the Fe $3d$
shell, at least on the static Hartree-Fock level, we calculated the
electronic structure of the LT phase of Fe$_3$O$_4$ using the LSDA+$U$
method.  Previous studies proved this method to be quite successful in
treating transition metal oxides with strong electron-electron correlations
as well as systems with long-range
order.\cite{AEH+96,AHA+01,AHY+03,ldau:LAZ95} The value of the $U$ parameter
for Fe cations estimated using different experimental and theoretical
technics lies in the range of 4.5-6 eV.\cite{AEH+96,ZSP+90,CHT+04} A
reasonably good agreement of the calculated gap value of 0.18 eV with the
experimental value \cite{PIT98} of 0.14 eV at 10 K was obtained using the
$U$ value of 5 eV.  Note, however, that the charge and orbital order
derived from the LSDA+$U$ calculations does not depend on the exact $U$
value when it is varied within the above mentioned limits. The value of the
Hund's coupling $J$=1 eV was estimated from constrained LDA
calculations.\cite{PEE+98} In the following all results presented in the
paper were obtained using a $U$ value of 5 eV.

\begin{figure}[tbp!]
\centerline{\includegraphics[height=0.45\textwidth,clip]{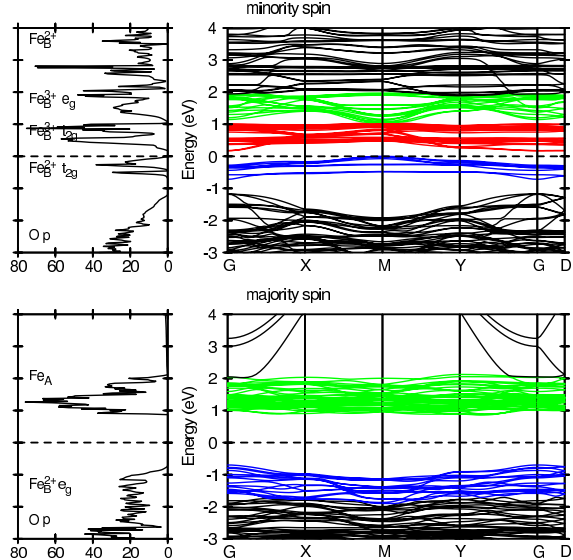}}
\caption{\label{fig:lsdaubnd} (Color online) Total DOS and band structure
of the $P2/c$ phase of Fe$_3$O$_4$ self-consistently obtained by the
LSDA+$U$ with $U$=5 eV and $J$=1 eV. The Fermi level is denoted by the
horizontal line and is taken as the zero of energy. A band gap of 0.18 eV
opens between M and $\Gamma$ symmetry points.  The energy bands
predominantly originating from the Fe$^{3+}_B$ $t_{2g}$ and $e_g$ states are
shown in dark gray (red) and light gray (green) colors, respectively,
whereas the gray (blue) color corresponds to Fe$^{2+}_B$ $t_{2g}$
bands. For the majority spin an energy gap of $\sim$2 eV opens between
Fe$^{2+}_B$ $e_g$ and Fe$_A$ bands shown in dark gray (red) and light gray
(green) color, respectively. The corresponding contributions to the total
DOS are shown in the left panel.}
\end{figure}

\begin{figure}[tbp!]
\centerline{\includegraphics[width=0.4\textwidth,clip]{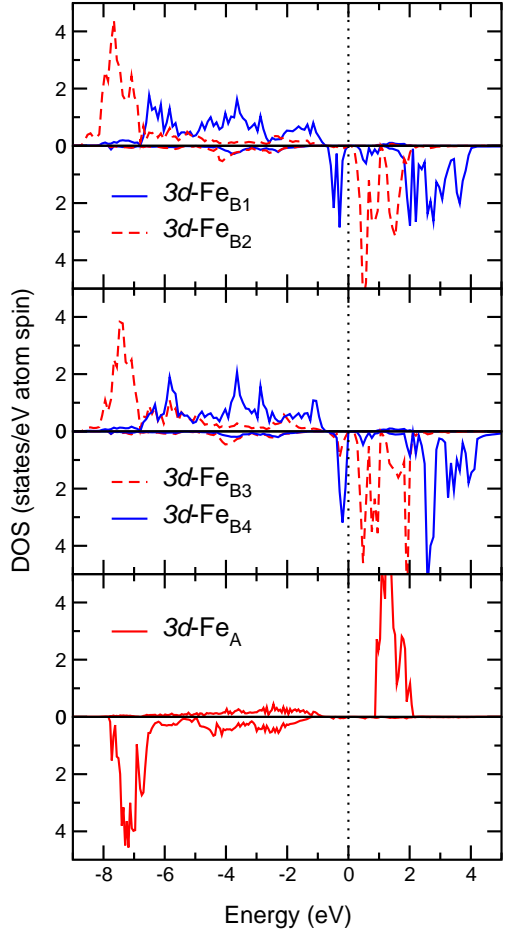}}
\caption{\label{fig:lsdaupdos} (Color online)
Partial DOS obtained from the LSDA+$U$ calculations with $U$=5 eV and 
$J$=1 eV for the $P2/c$ phase of Fe$_3$O$_4$. The Fermi level shown by 
dotted lines. A charge-ordered insulating solution is obtained. Fe $3d$ 
minority states corresponding to $B1$ and $B4$ sites are occupied (Fe$^{2+}$) 
and located just below the Fermi level, whereas 
$B2$ and $B3$ are empty (Fe$^{3+}$). The charge difference between 2+ and 3+ 
Fe$_B$ cations is found to be 0.23\,$\bar{e}$.}
\end{figure}

Figure~\ref{fig:lsdaubnd} shows the LSDA+$U$ band structure and the total
DOS calculated self-consistently for the low-temperature structure of
Fe$_3$O$_4$ using the Coulomb interaction parameter $U$=5 eV and exchange
coupling $J$=1 eV. The corresponding partial Fe$_B$ $3d$ DOS are shown in
Fig.~\ref{fig:lsdaupdos}. The LSDA+$U$ calculations give results qualitatively
distinct from those of the LSDA. An indirect energy gap of 0.18 eV opens in
the minority spin channel between M and $\Gamma$ symmetry points.  One of the
minority spin $t_{2g}$ states of Fe$_{B1}$ and Fe$_{B4}$ ions becomes
occupied while the Fe$_{B2}$ and Fe$_{B3}$ $t_{2g\downarrow}$ states are
pushed above the chemical potential. Although, as will be discussed
below, the calculated disproportion of Fe$_{B}$ $3d$ charges is
significantly less than 1, in the following we use the notations Fe$^{2+}$ and
Fe$^{3+}$ for Fe$_{B1,B4}$ and Fe$_{B2,B4}$ cations, respectively, having
in mind the difference of their $t_{2g}$ occupations.  The top of the
valence band is formed by the occupied $t_{2g\downarrow}$ states of $B1$ and
$B4$ Fe$^{2+}$ cations.  The bottom of the conduction band is formed
predominantly by the empty $t_{2g\downarrow}$ states of $B2$ and $B3$ Fe$^{3+}$
cations. The remaining unoccupied $t_{2g\downarrow}$ states of $B1$ and $B4$
Fe$^{2+}$ cations are pushed by the strong Coulomb repulsion to
energies above 2.5 eV.  Majority spin $3d$ Fe$_B$ states are shifted below
O $2p$ states, which form a wide band in the energy interval between $-$7
and $-2$ eV. This is in strong contrast to the uniform half-metallic
solution obtained by the LSDA.

Bands corresponding to the Fe$_A^{3+}$ cations are fully occupied (empty)
for minority (majority) spin states, respectively, and already in the LSDA
do not participate in the formation of bands near the Fermi level. The
LSDA+$U$ method does not strongly affect these bands, which lie in the
energy interval of $-$6 eV below and 1--2 eV above the Fermi level.

\subsection{Charge ordering}
\label{sec:co}

The obtained solution for CO of Fe$^{2+}$ and Fe$^{3+}$ cations on the $B$
sublattice is described by a dominant $[001]_c$ charge (and spin) density
wave, which originates from alternating chains of Fe$^{2+}$ ions on
octahedral $B1$ sites and Fe$^{3+}$ ions on $B2$ sites (see Table~I and
Fig.~2 in Ref.~\onlinecite{LYA04}).  A secondary $[00\frac{1}{2}]_c$
modulation in the phase of CO, which is formed by the chain of alternately
``occupied'' Fe$^{2+}$ ions on the $B4$ sites and ``empty'' Fe$^{3+}$ ions
on $B3$ sites, was found.  This is consistent with a $[001]$ nesting vector
instability at the Fermi surface in the Fe$_B$ minority electron states
which has been recently revealed by the LSDA calculations for the cubic
phase.\cite{YH99} The calculated CO scheme coincides with the class I CO
model proposed by Wright $et~al.$\cite{WAR01,WAR02}
All the tetrahedra formed by Fe$_B$ cations have either a 3:1 or 1:3 ratio of
Fe$^{2+}$ and Fe$^{3+}$ ions. Thus, the LSDA+$U$ calculations confirm that
the Anderson criterion is not satisfied in the LT phase.  However, it
should be pointed out that the Anderson criterion was introduced under the
assumption of equal interatomic distances within each tetrahedron, whereas
in the distorted LT structure the iron-iron distances vary from 2.86 to
3.05 \AA.  The same CO pattern has been recently confirmed by other LDA+$U$
calculations. \cite{JGH04}

An analysis of the $3d$ minority occupation matrices of Fe$_B$ cations
confirms very effective charge disproportion within the Fe$_B$ $t_{2g}$
minority spin subshell.  In particular, one of the $t_{2g \downarrow}$
states of Fe$^{2+}_{B1}$ and Fe$^{2+}_{B4}$ cations is almost completely
filled with the occupation $n\approx0.8$. On the other hand, the other two
$t_{2g \downarrow}$ orbitals of the Fe$^{2+}_{B}$ cations have
significantly smaller population of about 0.04.  The occupation numbers of
$t_{2g\downarrow}$ orbitals for Fe$^{3+}_{B2}$ and Fe$^{3+}_{B3}$ cations
do not exceed 0.1--0.17, which gives a value of about 0.7 for the largest
difference of the populations of Fe$^{2+}_{B}$ and Fe$^{3+}_{B}$
$t_{2g\downarrow}$ states. The occupation numbers of the minority spin
Fe$_{B}$ $3d$ orbitals and the net occupations of the $t_{2g\downarrow}$
and $e_{g \downarrow}$ states are given in the last two columns of Table
\ref{tab:lsdauorb}.

The change of the $t_{2g \downarrow}$ occupations caused by the charge
ordering is very effectively screened by the rearrangement of the other Fe
electrons. A significant contribution to the screening charge is provided by
Fe$_B$ $e_g$ states. Although the bands originating from these states are
located well above the energy gap, the minority spin $e_g$ orbitals form
relatively strong $\sigma$ bonds with 2$p$ states of the oxygen octahedron
and, as a result, give an appreciable contribution to the occupied part of the
valence band.  The energy of Fe$^{3+}_{B}$ $e_{g \downarrow}$ states is
lower than the energy of corresponding Fe$^{2+}_{B}$ states and the former
give a significantly larger contribution to the part of the valence band
formed mainly by O $2p$ states. Because of the stronger covalency of the
Fe$^{3+}_{B}$ $e_{g}$-- O $p$ bonds the net occupation of Fe$^{3+}_{B}$
$e_{g \downarrow}$ states becomes $\sim$0.25\,$\bar{e}$ larger (see the last
column of Table \ref{tab:lsdauorb}).  The resulting $3d$ charge difference
(0.23) and disproportionation of the total electron charges inside the
atomic spheres of Fe$^{2+}_{B}$ and Fe$^{3+}_{B}$ ions (0.24) are in
reasonably good agreement with the value of 0.2 estimated from a BVS
analysis of the $P2/c$ structure.\cite{WAR01,WAR02} The above-mentioned
screening of the changes in the Fe$_B$ $t_{2g}$ minority occupations
reduces the energy loss due to the development of charge order
incompatible with the Anderson criterion in the LT phase of Fe$_3$O$_4$.

Hence, due to the strong screening effects, the order parameter defined as
the difference of the net $3d$ charges of Fe$_B$ cations does not provide
conclusive evidence for CO.  This explains why the BVS analysis does not
give a convincing proof of CO existence. Apparently, a well-defined order
parameter is the difference of the occupations of the $t_{2g}$ minority
spin states for Fe$^{3+}_B$ and Fe$^{2+}_B$ cations which amounts to
70\% of the ideal ionic CO model and clearly indicates the existence of a
charge-ordered ground state below the Verwey transition.

The LSDA+$U$ calculations were also performed for the assumption of
Verwey charge order in the $P2/c$ structure. However, instead of the
assumed Verwey CO the same self-consistent solution as the one described
above was found. Therefore, the Verwey CO model is unstable in the
distorted $P2/c$ structure.  It is well known that with increasing
$U$ value localization is effectively increased. Remarkably, even for a
$U$ value increased up to 7--8 eV no Verwey-like CO pattern was found
self-consistently in the distorted $P2/c$ structure. On the contrary, the
LSDA+$U$ calculations performed for an undistorted $P2/c$ supercell of the
$Fd\bar{3}m$ structure result in an insulating CO solution which \textit{is
compatible} with the Verwey CO model.  Altogether this implies that the
Verwey CO model is unstable under a structure distortion from the
high-symmetry cubic into the low-symmetry $P2/c$ phase.

\begin{table}[tbp!]
\caption{\label{tab:lsdauorb} 
$3d$ orbital contribution to the formation of Fe$_B$ minority spin states
with occupancy $n$ evaluated by diagonalization of the occupation matrix.
Although one of the $t_{2g\downarrow}$ states of $B1$ and $B4$ sites is
almost occupied with $n\geq0.7$\,$\bar{e}$ the $t_{2g}$ minority spin
occupancies of $B2$ and $B3$ Fe$^{3+}$ cations are less than
0.1\,$\bar{e}$.  The occupied $t_{2g\downarrow}$ states of $B1$ and $B4$
Fe$^{2+}$ cations are predominantly of $d_{xz}\pm d_{yz}$ and $d_{x^2-y^2}$
character, respectively. The sum of $t_{2g}$ ($e_g$) occupations is given
in the last column.}
\begin{ruledtabular}
\begin{tabular}{lcccccccc}
Fe$_{B}$ ion & sym. & \multicolumn{3}{c}{$t_{2g}$} &\multicolumn{2}{c}{$e_{g}$}
&$n$ & $\sum_{t_{2g}}$\\
&          &   $yz$  &    $zx$ &$x^2-y^2$&$3z^2-r^2$&  $xy$  &      & 
($\sum_{e_{g}}$)\\ 
\hline
Fe$_{B1a}$ 
& $t_{2g}$ &   -0.47 &   -0.80 & \fm0.34 & \fm0.00 & -0.07   & 0.81 & 0.89 \\
&          & \fm0.86 &   -0.33 & \fm0.40 &   -0.01 & \fm0.01 & 0.04 & \\
&          &   -0.21 & \fm0.49 & \fm0.84 &   -0.01 & \fm0.04 & 0.04 & \\
& $e_{g}$  & \fm0.04 & \fm0.06 & \fm0.01 &   -0.49 & -0.87   & 0.15 & 0.26 \\
&          &   -0.02 &   -0.03 &   -0.02 &   -0.87 & \fm0.48 & 0.11 & \\
\hline
Fe$_{B1b}$ 
& $t_{2g}$ &   -0.47 & \fm0.83 & \fm0.28 &   -0.03 & \fm0.12 & 0.71 & 0.79 \\
&          & \fm0.83 & \fm0.31 & \fm0.46 & \fm0.00 & \fm0.00 & 0.04 & \\
&          & \fm0.30 & \fm0.45 &   -0.84 & \fm0.00 & \fm0.02 & 0.04 & \\
& $e_{g}$  &   -0.05 & \fm0.11 & \fm0.02 & \fm0.55 &   -0.83 & 0.15 & 0.27 \\
&          & \fm0.01 &   -0.04 & \fm0.00 & \fm0.83 & \fm0.55 & 0.12 & \\
\hline
Fe$_{B2a}$ 
& $t_{2g}$ & \fm0.00 & \fm0.91 & \fm0.41 & \fm0.00 & \fm0.00 & 0.09 & 0.24\\
&          &   -0.99 & \fm0.00 & \fm0.01 & \fm0.00 & \fm0.00 & 0.08 & \\
&          &   -0.01 & \fm0.41 &   -0.91 &   -0.04 & \fm0.00 & 0.07 & \\
& $e_{g}$  &   -0.01 & \fm0.00 & \fm0.00 & \fm0.00 & \fm0.99 & 0.27 & 0.52\\
&          & \fm0.00 &   -0.02 & \fm0.04 &   -0.99 & \fm0.00 & 0.25 & \\
\hline
Fe$_{B2b}$ 
& $t_{2g}$ & \fm0.00 &   -0.90 & \fm0.45 & \fm0.01 & \fm0.00 & 0.09 & 0.23\\
&          &   -0.99 &   -0.01 &   -0.03 & \fm0.00 &   -0.05 & 0.07 & \\
&          & \fm0.03 &   -0.45 &   -0.89 &   -0.04 & \fm0.00 & 0.07 & \\
& $e_{g}$  & \fm0.00 &   -0.01 &   -0.04 & \fm0.99 & \fm0.00 & 0.26 & 0.52 \\
&          &   -0.05 & \fm0.00 & \fm0.00 & \fm0.00 & \fm0.99 & 0.26 & \\
\hline
Fe$_{B3}$  
& $t_{2g}$ & \fm0.79 &   -0.18 & \fm0.53 &   -0.09 & \fm0.24 & 0.17 & 0.36 \\
&          &   -0.22 & \fm0.73 & \fm0.63 & \fm0.13 &   -0.10 & 0.11 & \\
&          & \fm0.52 & \fm0.65 &   -0.55 &   -0.08 &   -0.04 & 0.08 & \\
& $e_{g}$  &   -0.21 & \fm0.12 & \fm0.04 &   -0.91 & \fm0.35 & 0.25 & 0.49 \\
&          & \fm0.13 &   -0.11 & \fm0.11 &   -0.39 &   -0.90 & 0.24 & \\
\hline
Fe$_{B4}$  
& $t_{2g}$ &   -0.55 &   -0.18 & \fm0.82 &   -0.02 & \fm0.02 & 0.80 & 0.87\\
&          & \fm0.51 &   -0.85 & \fm0.16 & \fm0.01 &   -0.04 & 0.04 & \\
&          & \fm0.66 & \fm0.50 & \fm0.55 &   -0.03 & \fm0.01 & 0.03 &\\
& $e_{g}$  &   -0.01 &   -0.02 &   -0.03 &   -0.99 &   -0.12 & 0.13 & 0.25 \\
&          &   -0.02 & \fm0.03 & \fm0.02 & \fm0.12 &   -0.99 & 0.11 &\\
\end{tabular}
\end{ruledtabular}
\end{table}

Also we performed LSDA+$U$ calculations with the same $U$ and $J$ parameters 
(5 and 1 eV, respectively) for the assumption of one of the 32 class II CO 
models within $Cc$ supercell of $P2/c$, which is shown in Fig.~2 in 
Ref.~\onlinecite{WAR02}. But we found that this kind of CO is unstable 
and the self-consistent solution coincides with the one found 
previously for the $P2/c$ structure. 

Comparing the LSDA+$U$ results for the undistorted and distorted $P2/c$
unit cells we can conclude that the charge-ordering pattern of
Fe$^{2+}_{B}$ and Fe$^{3+}_{B}$ cations in the LT phase of Fe$_3$O$_4$,
derived from the BVS analysis in Ref.~\onlinecite{WAR02} and confirmed by
our study, \textit{is mainly forced} by the local distortions of the
crystal structure. Obviously we did not manage to study all possible charge-
ordering scenarios within $P2/c$ or $Cc$ supercell of $P2/c$. But our
results consistently indicate the importance of the small amplitude of
atomic displacements (almost of 0.07\AA) recently resolved by x-ray and
neutron powder diffraction.\cite{WAR01,WAR02} The additional displacements
leading to the $Cc$ supercell were estimated to be of $\sim$0.01 \AA\ but
have not been fully resolved so far. They also may be important for full
understanding of the CO in Fe$_3$O$_4$.  In particular, in the $P2/c$
subcell the true atomic positions are averaged over the corresponding
number of subsites in the $Cc$ cell. Therefore, the actual arrangement of
the locally Fe$_B$O$_6$ octahedra in the true $Cc$ structure can be more
complex, probably resulting in a more complicated charge and/or orbital
order for the LT structure.
The present calculations indicate that the competition of the ``elastic''
and electrostatic energy contributions in the total energy appears to be
responsible for the CO, which is realized in the LT structure of
Fe$_3$O$_4$.\cite{LYA04} Because of this, the Verwey CO model, which has the
lowest electrostatic but significant ``elastic'' energy contribution in the
total energy, becomes less favorable than other arrangements.

\subsection{Orbital ordering}

The self-consistent solution obtained by the LSDA+$U$ is not only charge
but also orbitally ordered. It is clearly seen from
Table~\ref{tab:lsdauorb}, which presents the contribution of $3d$ cubic
harmonics to the formation of Fe$_B$ minority spin states with an occupancy
$n$ (next to last column in Table~\ref{tab:lsdauorb}) evaluated by
diagonalization of the corresponding occupation matrix self-consistently
obtained by the LSDA+$U$.

As shown in the table the most occupied Fe$^{2+}$ $3d$ minority orbitals
are centered on the $B1a$, $B1b$, and $B4$ iron sites and have
$d_{xz}-d_{yz}$, $d_{xz}+d_{yz}$, and $d_{x^2-y^2}$ character,
respectively. Remarkably, the occupied $t_{2g \downarrow}$ orbitals of
Fe$_B$ cations are almost orthogonal to each other, i.e., their relative
orientation corresponds to an anti-ferro-orbital order. Since all Fe$_B$
cations are ferromagnetically coupled the obtained orbital order conforms
with the anti-ferro-orbital ferromagnetic state, which is the ground state
of the degenerate Hubbard model according to the Kugel-Khomskii
theory.\cite{KH75,KH82} This orbital order is consistent with the
corresponding distortions of FeO$_6$ octahedra. In particular, using simple
considerations which take into account only the change of the Fe-O bond
lengths and neglect the bending of the bonds, it was previously concluded
that the calculated orbital order is mainly determined by the distortions
of oxygen octahedra surrounding Fe$_{B}$ sites.\cite{LYA+05}

Also this simple analysis shows a remarkable difference between
Fe$_{B1}^{2+}$ and Fe$_{B4}^{2+}$ cations; namely, the average Fe$_{B1a}$-O
distance in the plane of occupied $d_{xz}-d_{yz}$ orbital is 2.087 \AA,
whereas in the planes of two other $t_{2g}$ orbitals they are only 2.063
 and 2.067 \AA. This difference between the average cation-anion
distance in the planes of occupied and unoccupied orbitals is remarkably
larger for Fe$_{B1a,b}$ (more than 0.02 \AA), although for Fe$_{B4}$ they
are 2.074 and 2.067 \AA\ for occupied $d_{x^2-y^2}$ and unoccupied
$d_{xz} \pm d_{yz}$ orbitals, respectively, which gives a difference of
only 0.007 \AA.  This small difference can be changed by applying a
uniaxial stress to the $P2/c$ unit cell resulting in modification of the 
electronic properties.\cite{H03} In particular, a few percent of
magnitude elongation of the $P2/c$ unit cell along the $c$ axis with
simultaneous (in order to preserve the same unit cell volume) compression
in the $ab$ plane, gives rise to orbital-order crossover on the
Fe$_{B4}$ site from a $d_{x^2-y^2}$ to a $d_{xz}$ occupied orbital.  At the
same time the charge order and occupied orbitals on the Fe$_{B1a,b}^{2+}$
sites remain the same. The pressure-induced spatial reorientation of the
occupied Fe$_{B4}$ $t_{2g}$ orbital was proved by the LSDA+$U$ calculations
for the strained $P2/c$ unit cell.  Note, however, that these rough
estimations do not take into account the elastic anisotropy in Fe$_3$O$_4$.
Moreover, the analysis was performed for the ``averaged'' $P2/c$
structure.  However, they provide insight into the orbital-ordering
phenomena behind the Verwey transition in magnetite as well as the problem
of an external parameter-controlled electron state (for example, orbital
ordering) in solids.\cite{LWS+05}

\subsection{Magnetic moments}
\label{sec:mom}

The strong variation of the occupancies of the minority spin Fe$_B$
$t_{2g}$ states leads to a pronounced modulation of the spin magnetic
moments on the $B$ sublattice. While the total moment per formula unit
remains at 4\,$\mu_{\text{B}}$ the magnetic moments of the Fe$^{2+}$ $B1$ 
(3.50\,$\mu_{\text{B}}$) 
and $B4$ (3.48\,$\mu_{\text{B}}$) cations become appreciably
smaller than Fe$_{B2}$ (3.94\,$\mu_{\text{B}}$) and Fe$_{B3}$ 
(3.81\,$\mu_{\text{B}}$) moments.
The $[001]_c$ charge and spin modulation on the $B$ sublattice is
accompanied by formation of a weak spin modulation on the oxygen ions
caused by different strengths of the hybridization of O $2p$ states with the
minority spin $3d$ states of Fe$^{2+}_{B}$ and Fe$^{3+}_{B}$ ions. In
particular, the oxygen magnetic moment reaches its maximal value of
$\sim$0.1\,$\mu_B$ for O3 and O4 ions, which lie in the plane of
Fe$^{3+}_{B2}$ cations. It substantially decreases for other oxygen ions and
approaches minimum for O1 and O2 lying in the plane of Fe$^{2+}_{B1}$
cations ($\sim$0.04\,$\mu_B$).

Recently, an anomalously large value of the Fe$_B$ orbital magnetic moment
reaching 0.33\,$\mu_{\mathrm{B}}$ has been deduced by applying sum rules to
experimental $L_{2,3}$ x-ray magnetic circular dichroism spectra of
Fe$_3$O$_4$.\cite{HCJG+04} In addition, the unquenched Fe$_B$ orbital
moment was also reported to be confirmed by the LDA+$U$
calculations. Later, however, this experimental finding was questioned by
Goering \textit{et al.}\cite{GGLS06} The average orbital moments between
$-$0.001\,$\mu_{\mathrm{B}}$ and 0.06\,$\mu_{\mathrm{B}}$ were found from
x-ray magnetic circular dichroism (XMCD) sum rules depending on the
integration range. From our spin-polarized relativistic LSDA+$U$
calculations for the LT structure we obtained the orbital moments of
0.19\,$\mu_{\mathrm{B}}$ and 0.014\,$\mu_{\mathrm{B}}$ for Fe$_{B1}$ and
Fe$_{B2}$ ions. Somewhat larger values of 0.039\,$\mu_{\mathrm{B}}$ and
0.22\,$\mu_{\mathrm{B}}$ were calculated for Fe$_{B3}$ and Fe$_{B4}$
cations, respectively. Taking into account the negative Fe$_A$ orbital
moment of $-$0.021\,$\mu_{\mathrm{B}}$ this gives the value of
0.07\,$\mu_{\mathrm{B}}$ for the average orbital moment. Thus, in agreement
with the previous theoretical results of Ref.\ \onlinecite{AHA+01} and XMCD
sum rule data of Ref.\ \onlinecite{GGLS06}, our calculations give the value
of Fe orbital moment of $\sim$0.07\,$\mu_{\mathrm{B}}$ which is much
smaller than reported in Ref.\ \onlinecite{HCJG+04}.

\subsection{Exchange coupling constants}
\label{sec:exchange}

\begin{table}[tbp!]
\caption{\label{tab:lsdauleip} 
Exchange couplings $J_{ij}$ (all with 
$|J_{ij}|>10$ K) are presented. The values are given in kelvin. 
The spatial representation 
of the Fe$_B$--Fe$_B$ exchange couplings is schematically 
shown in Fig.~\ref{fig:lsdauleip}.
$J_{ij}$ were calculated between the sublattices 
formed by the translations of the following Fe sites:
Fe$_{A1}$ (1/4,0.0034,0.06366), 
Fe$_{A2}$ (1/4,$-$0.4938,0.18867), 
Fe$_{B1a}$ (0,1/2,0), 
Fe$_{B1b}$ (1/2,1/2,0), 
Fe$_{B2a}$ (0,0.0096,1/4), 
Fe$_{B2b}$ (1/2,0.0096,1/4), 
Fe$_{B3}$ ($-$1/4,0.2659,0.1198),
Fe$_{B3}'$ (1/4,$-$0.2659,$-$0.1198), 
Fe$_{B3}''$ (1/4,0.2659,0.3801), 
Fe$_{B4}$ (1/4,0.2479,$-$0.1234), 
Fe$_{B4}'$ ($-$1/4,$-$0.2479,0.1234), and
Fe$_{B4}''$ (1/4,$-$0.2479,0.3765).
}
\begin{ruledtabular}
\begin{tabular}{llc}
$i$-atom & $j$-atom & $J_{ij}$, K \\ 
\hline
Fe$_{A1}$ & Fe$_{B1a}$ & $-$69.7  \\
          & Fe$_{B1b}$ & $-$69.9  \\
          & Fe$_{B2a}$ & $-$42.0  \\
          & Fe$_{B2b}$ & $-$42.1  \\
          & Fe$_{B3}$    & $-$73.7  \\
          & Fe$_{B3}'$   & $-$48.1  \\
          & Fe$_{B4}$    & $-$39.2  \\
          & Fe$_{B4}'$   & $-$52.9  \\
\hline
Fe$_{A2}$ & Fe$_{B1a}$ & $-$40.0  \\
          & Fe$_{B1b}$ & $-$27.0  \\
          & Fe$_{B2a}$ & $-$74.8  \\
          & Fe$_{B2b}$ & $-$77.4  \\
          & Fe$_{B3}$    & $-$89.1  \\
          & Fe$_{B3}''$  & $-$31.1  \\
          & Fe$_{B4}'$   & $-$62.7  \\
          & Fe$_{B4}''$  & $-$27.8  \\
\hline\hline
Fe$_{B1a}$ & Fe$_{B3}$, Fe$_{B3}'$ & +12.7  \\
Fe$_{B1b}$ & Fe$_{B3}$, Fe$_{B3}'$ & +27.5  \\
Fe$_{B2a}$ & Fe$_{B2b}$ & $-$11.6  \\ 
Fe$_{B3}$    & Fe$_{B4}'$ & +16.8  \\ 
Fe$_{B4}$    & Fe$_{B3}'$ & +16.8  \\
\end{tabular}
\end{ruledtabular}
\end{table}

\begin{figure}[tbh!]
\centerline{\includegraphics[width=0.45\textwidth,clip]{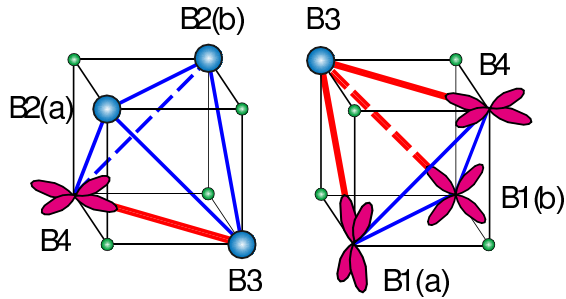}}
\caption{ (Color online) Sketch of the spatial arrangement of exchange
interaction parameters between the octahedral Fe$_B$ sites. Orbitals
approximate the occupied $3d$ minority orbitals of Fe$_B^{2+}$
cations. Fe$_B^{3+}$ cations are shown by large (blue) spheres.  Oxygen atoms
are shown by small (green) spheres. Ferromagnetic couplings between Fe$_B$
cations are shown by the thick (red) lines, whereas antiferromagnetic exchanges
depicted  by the thin (blue) lines.}
\label{fig:lsdauleip}
\end{figure}

To proceed further we performed calculations of the exchange interaction
parameters $J_{ij}$ via the variation of the ground state energy with
respect to the magnetic moment rotation angle.\cite{ldau:LAZ95} The
exchange coupling parameter $J_{ij}$ represents the effective pair exchange
interaction between the $i$th and $j$th Fe atoms with effective Heisenberg
Hamiltonian $H = - \sum_{i > j} J_{ij} {\bf S}_i \cdot {\bf S}_j$. Here,
${\bf S}_i$ and ${\bf S}_j$ are the spins at site $i$ and
$j$ (5/2 and 2 for Fe$^{3+}$ and Fe$^{2+}$ cations,
respectively). Positive (negative) values of $J_{ij}$ correspond to the
ferromagnetic (antiferromagnetic) coupling between sites.  As shown in
Table~\ref{tab:lsdauleip} the exchange couplings between $A$ and $B$ iron
sublattices are rather large, of about $-$70 K, and antiferromagnetic. 
The Fe$_A$--Fe$_A$ interactions are weakly antiferromagnetic with
the maximal absolute value of 9.3 K (not shown in
Table~\ref{tab:lsdauleip}). The exchange couplings between the Fe$_B$
sites ($|J_{BB}| \leq 27.5$ K) are substantially smaller than
Fe$_A$--Fe$_B$ ones and almost all of them are ferromagnetic 
(see Fig.~\ref{fig:lsdauleip}). 
Weak antiferromagnetic
couplings with $|J_{BB}| \leq 11.6$ K are also obtained [mainly between
the sites with the same 2+ or 3+ valence state, shown by the thin (blue)  
lines in Fig.~\ref{fig:lsdauleip}]. The spatial
representation of these exchange couplings is presented in
Fig.~\ref{fig:lsdauleip}. Other couplings that are not shown in 
Table~\ref{tab:lsdauleip} are weaker than 10 K.

Experimental estimation of the exchange couplings in Fe$_3$O$_4$ was
first performed by N\'eel on the basis of the two-sublattice collinear
model.\cite{N48} From analysis of the temperature behaviour of the
saturation magnetization and paramagnetic susceptibility he obtained
$J_{AA}=-17.7$, $J_{AB}=-23.4$, and $J_{BB}=0.5$ K, where $A$ and $B$
refer to the tetrahedral and octahedral Fe sites, respectively.  We find
these values qualitatively in accord with our results presented in
Table~\ref{tab:lsdauleip}; namely, as in N\'eel's model, the
calculations result in strong antiferromagnetic coupling between the $A$
and $B$ sublattices; $J_{AA}$ couplings (not shown in
Table~\ref{tab:lsdauleip}) are considerably smaller than $J_{AB}$; the
exchange couplings in the $B$ sublattice are weak and almost all of them
are ferromagnetic. On the other hand, the small antiferromagnetic
Fe$_{B2a}^{3+}$--Fe$_{B2b}^{3+}$ exchange interaction (see
Table~\ref{tab:lsdauleip}) is in exact agreement with recent estimations
using the two-sublattice model.\cite{LN02} Three-sublattice model
calculations give an overall similar result, except, however, the
Fe$_B^{2+}$--Fe$_B^{2+}$ exchange coupling, which seems to be
overestimated.\cite{SSN79}

\section{Spectral properties}
\label{sec:optics}

\subsection{XPS spectra}
\label{sec:xps}

\begin{figure}[tbp!]
\centerline{\includegraphics[width=0.45\textwidth,clip]{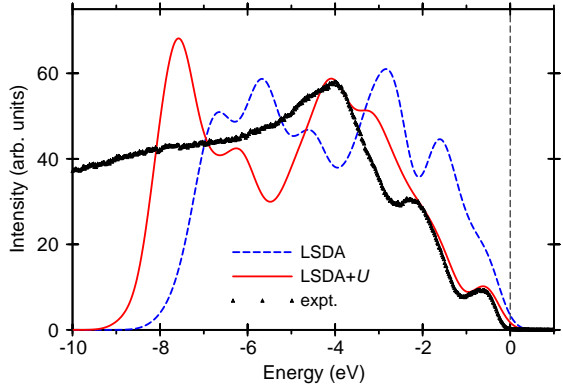}}
\caption{\label{fig:xps} (Color online) Comparison of the total DOS
obtained from LSDA (dashed line) and LSDA+$U$ calculations with $U$=5 eV
and $J$=1 eV (solid line) to the experimental XPS spectrum taken from Ref.\
\onlinecite{SST05} (triangles). A rigid shift of $-$0.3 eV is applied to
the calculated DOS curves.}
\end{figure}

In Fig.\ \ref{fig:xps} the LSDA and LSDA+$U$ total DOSs are compared to the
soft x-ray photoemission (XPS) spectrum from Ref.\ \onlinecite{SST05}. No
attempts to account for different cross sections of Fe $3d$ and O $2p$
states have been made. To account for the experimental resolution and
finite lifetime broadening the theoretical spectrum was convoluted with a
Gaussian with a half-width of 0.4 eV. The calculated DOS curves had to be
shifted by 0.3 eV to lower energies in order to align the experimental
peak at $\sim-0.7$ eV with the corresponding peak in the LSDA+$U$ DOS.
The necessity of the rigid shift can be attributed to the problem of
experimental determination of the Fermi level position for insulating
samples. A similar shift of 0.2 eV was used in Ref.\ \onlinecite{CRWB98} when
comparing photoelectron spectra to LSDA calculations.
The experimental spectrum was measured at 100 K using Fe $2p$--$3d$
resonance photoemission with the photon energy $h\nu=707.6$ eV. Because of
the the relatively large escape depth of photoelectrons at this photon
energy it is expected that the spectrum mainly probes bulk electronic
states of Fe$_3$O$_4$.

Figure \ref{fig:xps} shows that the shift of the position of the occupied
majority spin Fe$_B$ and minority spin Fe$_A$ states to lower energies in
the LSDA+$U$ calculation greatly improves the agreement with the
experiment. A feature at $\sim-$0.7 eV, which transforms to a well defined
peak in the LSDA+$U$ calculations, can be identified as $d^6\rightarrow
d^5$ transitions at Fe$_B^{2+}$ ions. This peak is completely spin
polarized in accordance with the finding of a recent spin-resolved
photoemission study.\cite{HCCT+02}
From a comparison with the $l$-projected DOS shown in Fig.\ \ref{fig:lsdaupdos}
one can associate an experimental peak at 2 eV with transitions from the
majority spin Fe$_B^{2+}$ $e_g$ states which give the predominant contribution
in this energy range. In the calculated DOS, however, this feature is
masked by a peak at $\sim-$3 eV and appears only as a shoulder. 
A peak at 4 eV in the experimental spectrum originates to a great extent from
the transitions from the majority spin Fe$_B^{2+}$ $t_{2g}$ strongly
hybridized with O $2p$ states. Note, however, that $3d$ states of other Fe
ions as well as O $2p$ states also contribute to the total DOS in the
energy range from $-7$ to $-4$ eV. Finally, a DOS peak at $-7.8$ eV formed
mainly by the majority spin Fe$_B^{3+}$ and minority spin Fe$_A$ $3d$ states
accounts for a weak feature found at $-8$ eV in the experimental
spectrum. This feature seems to be masked by a high-energy satellite but
can be clearly seen in Fe $3p$--$3d$ resonance photoemission 
($h\nu=56$ eV).\cite{CRWB98}

\subsection{Optical and magneto-optical spectra}
\label{sec:omo}

\begin{figure}[tbp!]
\centerline{\includegraphics[width=0.45\textwidth,clip]{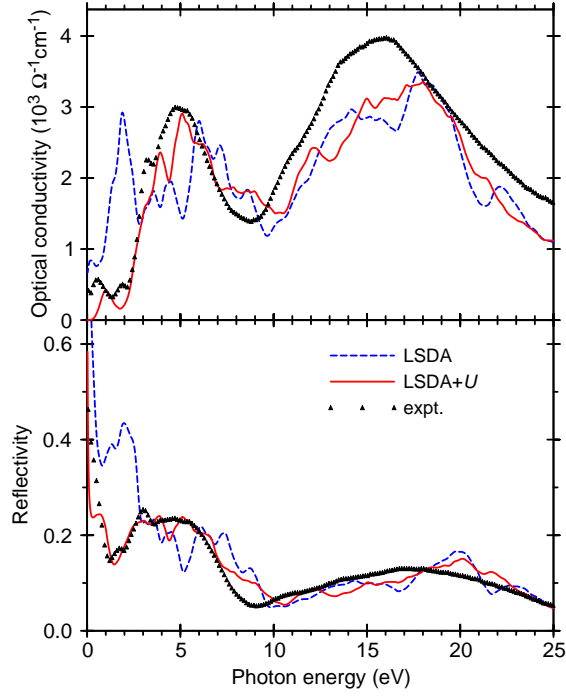}}
\caption{\label{fig:sigrw} (Color online) Reflectivity (lower panel) and
optical conductivity (upper panel) spectra calculated using the LSDA
(dashed line) and LSDA+$U$ with $U$=5 eV and $J$=1 eV (solid line) for the
$P2/c$ model of the LT phase of Fe$_3$O$_4$. Room temperature experimental
spectra from Ref.\ \onlinecite{PIT98} are shown by triangles.}
\end{figure}

Motivated by the fact that previous calculations were performed for the
undistorted {\it cubic} structure assuming the Verwey CO
pattern,\cite{AHA+01} in this section we present the optical and
magneto-optical spectra calculated for the $P2/c$ structural model of the
LT phase of Fe$_3$O$_4$.  Figure \ref{fig:sigrw} shows optical conductivity
(top) and reflectivity (bottom) calculated using the LSDA and LSDA+$U$ with
$U$=5 eV and $J$=1 eV.  Symmetry lowering from the fcc to $P2/c$ structure
causes some degree of optical anisotropy in the low-frequency
range. However, it is rather weak and here we present the spectra averaged
over different photon polarizations. The results are compared to the
experimental optical spectra taken from Ref.~\onlinecite{PIT98}, which
reveal strong temperature dependence only in the narrow energy range of
0--1 eV.

The LT structure distortions have a rather weak effect on the LSDA spectra
which resemble closely the spectra calculated for the cubic
phase.\cite{AHA+01} This finding agrees well with the fact that the LSDA
band structure presented above is not strongly affected by the crystal
structure distortions. It is worth recalling that the LSDA calculations
result in a metallic solution and therefore give the wrong asymptotic
behavior for the optical reflectivity at $\omega\rightarrow 0$.  Moreover,
a peak at $\sim$6.5 eV in the calculated conductivity spectrum is shifted
to higher energies with respect to the experimental peak centered at
$\sim$5 eV. Both conductivity and reflectivity show peaks at 2 eV which
are much higher than in the experimental spectra. Hence, the LSDA theory
fails to correctly reproduce the low-energy spectral properties of
Fe$_3$O$_4$, whereas it describes reasonably well the experimental spectra
above 10 eV.

\begin{figure}[tbp!]
\centerline{\includegraphics[width=0.45\textwidth,clip]{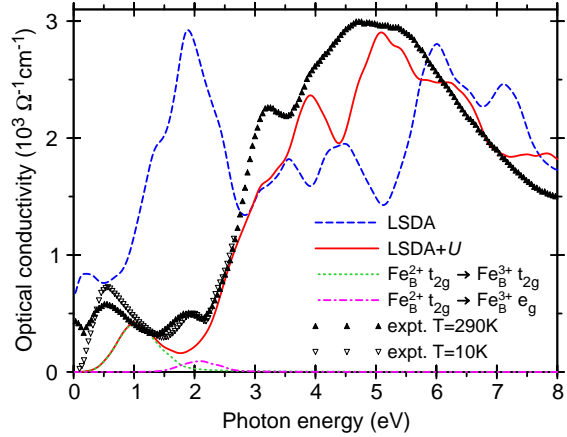}}
\caption{\label{fig:sig} (Color online) Comparison of LSDA (dashed line)
and LSDA+$U$ with $U$=5 eV and $J$=1 eV (solid line) optical conductivity
spectra to the experimental spectra measured at room temperature (up
triangles) and at 10 K (down triangles). \cite{PIT98} Also plotted are the
contributions of the interband transitions from occupied minority spin
Fe$_B^{2+}$ $t_{2g}$ to Fe$_B^{3+}$ $t_{2g}$ (short-dashed line) and
Fe$_B^{3+}$ $e_{g}$ (dash-dotted line) to the LSDA+$U$ spectrum.}
\end{figure}

As in Ref.~\onlinecite{AHA+01} better agreement between the calculated
and experimental spectra was obtained using the LSDA+$U$ method.  Since the
self-consistent LSDA+$U$ solution is insulating the reflectivity at zero
frequency is no longer equal to 1 but approaches the value of 0.6 which,
however, is somewhat higher than in the experiment.  Figure \ref{fig:sig}
shows the expanded view of the theoretical and experimental\cite{PIT98}
optical spectra in a smaller frequency range below 8 eV.  Charge and
orbital ordering within the minority spin Fe$_B$ $t_{2g}$ states changes
drastically the spectrum below 3 eV.  Because of the shift of the
unoccupied minority spin Fe$_B^{2+}$ and majority spin Fe$_A$ $d$ states to
higher energies as compared to their LSDA positions the spectral weight is
transferred from 2 eV to higher frequencies. This improves the agreement
with the experimental optical conductivity in the range 3.5--6 eV.
However, instead of the two-peak structure at 0.5 and 2 eV the LSDA+$U$
calculations reveal a maximum at $\sim$1 eV which can be associated with
the 0.5 eV peak in the experimental spectrum. The theoretical peak is,
however, more symmetric, shifted to higher frequencies, and its magnitude
is lower than in the experiment.  An analysis of different interband
contributions to the calculated spectra shows that the 1 eV peak is
predominantly formed by the transitions from occupied Fe$_B^{2+}$
$t_{2g\downarrow}$ into empty Fe$_B^{3+}$ $t_{2g\downarrow}$ bands. The
maximum of the Fe$_B^{2+}$ $t_{2g\downarrow}$ to Fe$_B^{3+}$
$e_{g\downarrow}$ interband conductivity coincides with the second
experimental peak at $\sim$2 eV. However, its theoretical intensity is too
low and these transitions do not manifest themselves as a peak in the
spectrum.

An attempt to make the agreement between the theoretical and experimental
spectra better by decreasing the value of $U$ has not given any substantial
improvements.  In particular for $U$=4.5 eV (not shown here) the 1 eV
peak shifts to lower energies only by 0.2 eV. Moreover, its shape is
still more symmetric than the shape of the corresponding peak in the
experimental spectra at 10 K,\cite{PIT98} whereas the theoretical
intensity of the second peak remains too low.

A possible reason for the abovementioned discrepancies between the
theoretical and experimental spectra below 2 eV can be neglecting
dynamical correlations in the frame of the static LSDA+$U$ method.
Moreover, it is worth recalling that the true LT crystal structure has not
been fully resolved yet and the calculations were performed for the
``averaged'' $P2/c$ structural model for the LT phase of Fe$_3$O$_4$.  As
described in Sec.\ \ref{sec:co} the self-consistently obtained CO
corresponds to the class I CO model.\cite{WAR02} If one of the class II CO
patterns is realized in the true $Cc$ structure, as has been suggested by
a recent resonant x-ray diffraction study, \cite{GWAR05} then more
complicated charge and orbital order may result in narrowing of Fe$_B$ 
$d$-derived bands and a shift of the peak in optical absorption to lower
energies.  Therefore, further investigations of the LT crystal structure
are strongly demanded.

\begin{figure}[tbp!]
\centerline{\includegraphics[width=0.45\textwidth,clip]{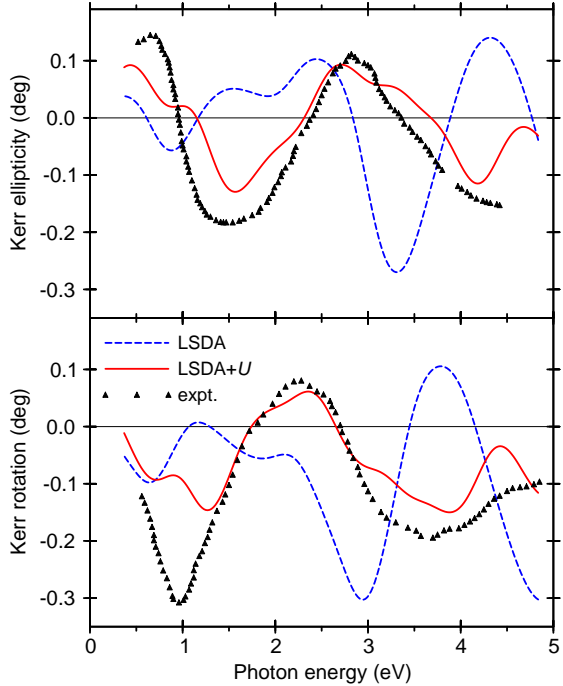}}
\caption{\label{fig:kerr} (Color online) Polar Kerr rotation (lower panel)
and ellipticity (upper panel) calculated for the $P2/c$ model of the LT
phase of Fe$_3$O$_4$ using LSDA (dashed line) and LSDA+$U$ with $U$=5 eV
and $J$=1 eV (solid line). Experimental spectra measured at room
temperature for an annealed synthetic sample of Fe$_3$O$_4$ in Ref.\
\onlinecite{ZSW81} are plotted by triangles.}
\end{figure}

In Fig.\ \ref{fig:kerr} magneto-optical polar Kerr rotation and ellipticity
spectra calculated for the $P2/c$ structural model are compared to the room
temperature spectra of the annealed synthetic sample of
Fe$_3$O$_4$.\cite{ZSW81} Calculations were performed assuming that the
magnetization and light propagation directions are parallel to the $c$ axis
of the $P/2c$ cell. Similar to the optical conductivity spectra the use of
the LSDA+$U$ approach improves the agreement with the experiment above 2 eV. 
The minimum at 1.3 eV in the theoretical Kerr rotation spectrum is,
however, less pronounced and shifted to higher frequencies as compared to
the experimental one (measured at room temperature).  Comparison of the
magneto-optical spectra for the LT phase to the spectra calculated assuming
the Verwey CO in the fcc unit cell \cite{AHA+01} shows that the
theoretical spectra are rather sensitive to the structural and CO model
used in the calculations. Its experimental measurements for the LT phase
of Fe$_3$O$_4$ and comparison with the theoretical prediction presented in
Fig.~\ref{fig:kerr} are of strong importance.

\subsection{O $K$-edge XAS}
\label{sec:oxas}

\begin{figure}[tbp!]
\centerline{\includegraphics[width=0.45\textwidth,clip]{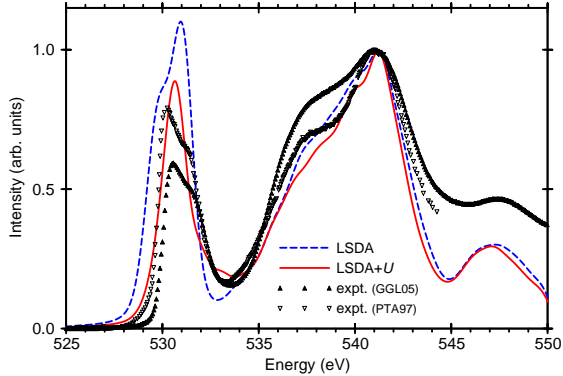}}
\caption{\label{fig:OKxas} (Color online) LSDA (dashed line) and LSDA+$U$
with $U$=5 eV and $J$=1 eV (solid line) O $K$-edge x-ray absorption spectra
calculated for the $P2/c$ model of the LT phase of Fe$_3$O$_4$. The sum of
the spectra from inequivalent O sites is shown. Experimental spectra from
Refs.\ \onlinecite{PTA97} and \onlinecite{GGLS+05} are plotted by down and
up triangles, respectively.}
\end{figure}

In the process of x-ray absorption at the O $K$ edge a core O 1$s$ electron
is excited into unoccupied states in the conduction band. In the dipole
approximation, which was used in the present calculations, the final states
for the transition are of $p$ symmetry and the spectrum probes the density
of unoccupied O 2$p$ states. O $K$-edge x-ray absorption spectra calculated
using the LSDA and LSDA+$U$ (with $U$=5 eV and $J$=1 eV) and averaged over
eight inequivalent O sites in the $P2/c$ unit cell are presented in Fig.\
\ref{fig:OKxas}.  Experimental spectra taken from Refs.~\onlinecite{PTA97}
and \onlinecite{GGLS+05} are shown in Fig.~\ref{fig:OKxas} by down and up
triangles, respectively.  In order to account for the effect of the final
life time of the O 1$s$ core hole and experimental resolution the spectra
were broadened by convolution with Lorentzian and Gaussian functions of the
width of 0.4 and of 0.5 eV, respectively. Then the calculated spectra
were aligned and normalized to the main absorption peak at 541 eV in the
experimental spectrum.

As mentioned in Sec.\ \ref{sec:lsda} the lattice distortions themselves
have only a weak effect on the LSDA band structure of Fe$_3$O$_4$ and the LSDA
spectra calculated for inequivalent O sites in the $P2/c$ cell resemble
closely the O $K$-edge spectrum for the undistorted $Fd\bar{3}m$ structure.
Both the LSDA and LSDA+$U$ spectra reproduce well the shape of the 541 eV
peak and the position of a peak at 547.5 eV. Our calculations confirm the
conclusion of Ref.\ \onlinecite{WGJP+97} that these peaks originate from
transitions to bands formed mainly by delocalized Fe 4$s$ and 4$p$
states.

The so-called preedge peak at 531 eV is formed by transitions to O
2$p$ states hybridized with Fe$_A$ and Fe$_B$ $3d$ states and, thus, is
more sensitive to the changes of the electronic structure caused by charge
and orbital ordering effects.  The preedge peak in the LSDA spectrum is
wider than the experimental one, more intense, and shifted to lower
energies.  The experimental peak exhibits a second feature that appears as
a shoulder at the high-energy slope of the peak. The LSDA spectrum also has
a two-feature structure but their intensities are reversed. The
two-feature structure of the preedge peak was interpreted as coming from
the hybridization of O 2$p$ states with Fe $3d$ states split by the ligand
field. \cite{WGJP+97} Indeed, the comparison of the O $K$-edge spectrum
calculated for the undistorted fcc structure to the corresponding DOS
shows that the low-energy shoulder is formed by transitions to O 2$p$
states hybridized with majority spin Fe$_A$ $3d$ states of $e$ symmetry and
minority spin Fe$_B$ $t_{2g}$ states, whereas the main peak reflects the O
2$p$ hybridization with Fe$_A$ $t_{2\uparrow}$ and Fe$_B$ $e_{g\downarrow}$
states.  In the latter case the O $2p$ -- Fe $3d$ hybridization is stronger
and this causes the higher intensity of the corresponding feature in the
LSDA spectrum.  It should be noted that calculations of the O $K$-edge spectrum
based on the multiple-scattering approach reproduced the relative
intensities of the two features of the preedge peak. \cite{WGJP+97}

O $K$-edge spectra calculated for inequivalent O sites in the $P2/c$ unit
cell using the LSDA+$U$ approach show much stronger variation of the
shape of the preedge peak as compared to the LSDA calculations. The
strongest deviation of the peak shape is found for those O ions that have
only Fe$_B^{2+}$ cations among their nearest neighbors. In spite of that,
the preedge peak in the averaged spectrum becomes narrower than the LSDA
one and shifts to higher energies. An additional feature does appear at the
high-energy slope of the peak but its intensity is much weaker than in the
experiment.

We can conclude that LSDA+$U$ calculations reproduce better the energy
position of the preedge peak in the O $K$-edge spectrum but the shape of
the peak still differs from the experimentally observed one. Calculations
performed for other CO models possible in the true $Cc$ structure might in
principle improve the agreement with the experiment. We have to note,
however, that recent measurements of the O $K$-edge spectrum performed in a
wide temperature range below and above the Verwey transition show very
small variation of the shape of the preedge peak. \cite{GGLS+05} This
implies that the O $K$-edge spectrum should not be very sensitive to the
details of a particular CO pattern. Many-body effects that are beyond the
LSDA+$U$ may be responsible for the disagreement between the
theory and the experiment.

\section{Summary and conclusions}
\label{sec:sum}

In summary, in the present LSDA+$U$ study of the $P2/c$ model of the LT
phase of Fe$_3$O$_4$ we found a charge- and orbitally ordered insulator
with an energy gap of 0.18 eV. The obtained charge-ordered ground state is
described by a dominant $[001]_c$ charge density wave with a minor
$[00\frac{1}{2}]_c$ modulation on the Fe$_B$ sublattice. A weak $[001]_c$
spin/charge modulation on the oxygen ions was also obtained. The CO
coincides with the earlier proposed class I CO \cite{WAR01,WAR02} and
confirms violation of the Anderson criterion.\cite{And56} While the
screening of the charge disproportion is so effective that the total $3d$
charge disproportion is rather small (0.23), the charge order is well
pronounced with an order parameter defined as the difference of $t_{2g
\downarrow}$ occupancies of 2+ and 3+ Fe$_B$ cations (0.7). This agrees
well with the result of BVS analysis for a monoclinic structure (0.2). The
orbital order is in agreement with the Kugel-Khomskii theory \cite{KH75}
and corresponds to the local distortions of oxygen octahedra surrounding
Fe$_{B}$ sites. The average Fe orbital moment of $\sim0.07\,\mu_B$ agrees
well with the recent experimental findings.\cite{GLG06}

Calculations of the effective exchange coupling constants between Fe spin
magnetic moments show that the dominating interaction is an
antiferromagnetic coupling between Fe$_A$ and Fe$_B$ moments. The coupling
between Fe$_B^{2+}$ and Fe$_B^{3+}$ moments is found to be weaker and
ferromagnetic.

The relevance of the charge-ordered LSDA+$U$ solution has been verified by
performing calculations of optical, magneto-optical polar Kerr rotation,
and O $K$-edge x-ray absorption spectra and comparing them to available
experimental data. The LSDA+$U$ spectra show much better agreement with the
experimental ones as compared to the spectra calculated for the
charge-uniform LSDA solution.  Unfortunately, the remaining discrepancies
between the theory and the experiment in the low-frequency part of the
optical and MO spectra and in the shape of the preedge peak in x-ray
absorption do not allow us to discriminate among different CO models in the
true $Cc$ structure.

\begin{acknowledgments}
It is a pleasure to thank D.~Vollhardt, P.~Fulde, J.~P.~Attfield,
R.~Claessen, D.~Schrupp, A.~Pimenov, M.~A.~Korotin, and D.~I.~Khomskii for
helpful discussions. Also we wish to thank R.~Claessen and D.~Schrupp for
providing us experimental data.  The present work was supported by RFFI
Grants No.\ 04-02-16096, No.\ 03-0239024, No.\ 06-02-81017, and by DFG
through Grant No.\ 484.
\end{acknowledgments}


\end{document}